\documentclass[prb,twocolumn,showpacs,amsmath,amssymb,floats]{revtex4}
\usepackage[dvips]{graphicx}
\usepackage{color}

\def\be{\begin{equation}}
\def\ee{\end{equation}}
\def\bea{\begin{eqnarray}}
\def\eea{\end{eqnarray}}

\begin{document}
\title{Janssen effect and the stability of quasi 2-D sandpiles}
\author{Fatemeh Ebrahimi}
\author{Tahereh Azizpour}
\author{Hamed Maleki$^1$}
\affiliation{Physics Department, University of Birjand, Birjand
97175-615, Iran\\}
\date{\today}

\begin{abstract}
We present the results of  three dimensional molecular
dynamics study of global normal stresses in quasi two dimensional
sandpiles formed by pouring  mono dispersed cohesionless spherical
grains into a vertical granular Hele-Shaw cell. We observe Janssen
effect which is the phenomenon of pressure saturation at the
bottom of the container. Simulation of cells with different
thicknesses shows that the Janssen coefficient $\kappa$ is a
function of the cell thickness. Dependence of global normal
stresses as well as $\kappa$ on the friction coefficients between
the grains ($\mu_p$) and with walls ( $\mu_w$) are also studied.
The results show that in the range of our simulations $\kappa$
usually increases with wall-grain friction coefficient. Meanwhile
by increasing $\mu_p$ while the other system parameters are fixed,
we witness a gradual increase in $\kappa$ to a parameter dependent
maximal value.

\end{abstract}
\maketitle

\footnotetext[1]{present address: Department of Physics,
Johannes-Gutenberg University of Mainz, 55099 Mainz, Germany}

\section{Introduction}

Pouring a large amount of granular material into a vertical and
static cell produces two different, interesting structures
depending on the geometrical boundary conditions imposed on the
system. In a silo when all the side walls are closed, injection of
granular material produces a static granular pack which is a
vertical column with a flat and horizontal free surface. Granular
packs have been the subject of studies for long
time~\cite{jan,landry,sil,maks,mueth,blair,wolf} because, in
addition to the practical applications like corn storages in
silos, they serve as a useful model for a variety of physical
systems. On the other hand, inserting grains into a cell with at
least one open boundary, generates a stable heap with finite
slope. This non-zero angle to the horizontal, the angle of repose
$\theta$, is one of the most important characteristic of the
granular material at macroscopic
scales~\cite{bag,gh,bol,xy1,xy2,cou,maleki,maks2,maks3}.

A remarkable feature of  deep granular columns is the saturation
of vertical stress  at the bottom of the silo known as the Janssen
effect, named after German engineer H.A. Janssen~\cite{jan}.
Janssen was one of the first  who studied systematically the
effect of pressure screening at the bottom plate of silos and
succeeded to present a theoretical framework which accounts for
this peculiar behavior of granular matter. In Janssen's model
the granular material is treated as a continuous medium  in which
a fraction $\kappa$ of the vertical stress $\sigma_{zz}$ converts
to the horizontal stress. With the further assumption that all the
frictional forces exerted by the walls are at the Coulomb failure
criterion, he derived the following expression for the value of
$\sigma_{zz}$ at the depth $z$ of a granular pack in a vertical
container:
\begin{equation}
\sigma_{zz}(z)=\rho g l(1-e^{-\frac{z}{l}})
\end{equation}
where $\rho $ is the mass density of the granular material, $g$ is
the gravitational acceleration, and  $l$ is  the decay length. The
decay length (and subsequently the saturated stress
$\sigma_{zz}^{max}=\rho g l$) is related to the cell geometry and
wall-grain interaction as well as the grain-grain interactions.
Janssen's theoretical analysis leads to a value of
$l=\frac{A_b}{U\kappa \mu_w}$ for decay length in which $A_b$ and
$U$ are the area and circumference of the cell profile
respectively and $\mu_w$ is the friction coefficient between the
walls and grains (sometimes and especially in old literature, the
value of $K= \kappa \mu_w$ has been referred to as Janssen's
coefficient). Recent experimental and computational studies
revealed that the above formula needs  to be modified,  and as
such, the classical Janssen analysis and specifically the
assumption of Coulomb failure is not always satisfied. A number of
modified Janssen formulas (which usually mandate a region of
perfect hydrostaticity next to the surface) have been suggested
which fit the data obtained from real experiments and
sophisticated simulations~\cite{landry,ned,vanel,ovarlez} better
than the standard Janssen's formula. However, the concept of
pressure saturation at the bottom of the container is still valid.

The Janssen effect originates from the ability of granular materials
to support shear stress. The frictional forces exerted by
container's lateral walls carry part of the grains weight. These
very frictional forces are also responsible for the formation and
stability of a sandpile in a container with an open boundary. At
the beginning, when the  injected grains fall on the bottom of the
cell, the frictional forces between the grains and the cell walls
dissipate the major part of the grains kinetic energy, helping
the rapid formation of a sandpile. Meanwhile, the presence of
lateral walls modifies the angle of repose mostly because of the
arching phenomena between front and rear walls~\cite{gh}. In fact,
in the case of a thin container (usually called the granular
Hele-Shaw cell) many studies have demonstrated that just like the
saturated normal stress, the angle of repose $\theta$ is dependent
on both cell geometry (here the cell thickness $w$) and the
wall-grain friction coefficient~\cite{gh,bol,xy1,xy2,cou,maleki}.
Courrech et al ~\cite{cou} proposed a simple model to incorporate
the lateral wall effect on piling via the Janssen effect. This
model introduces a new characteristic length defined as $
B_r=2\kappa \mu _w h_{freeze} $, where  $h_{freeze}$ is the
flowing layer height when an induced surface avalanche  is about
to stop. The predicted growth of $\theta$ in Courrech et al's
model:
\begin{equation}
\frac{sin(\theta_ w)-sin(\theta_\infty)}{cos(\theta_ w)}
=1-\frac{w}{B_r}(1-e^{-\frac{B_r}{w}})
\end{equation}
is slower than the empirical exponential
law~\cite{gh}:
\begin{equation}
\theta_w=\theta_\infty(1+\alpha\: e^{-w/\Delta})
\end{equation}
which  has been traditionally used to fit their data.
Here, $\alpha$ is a constant which depends on grain properties and
$\Delta$ is a characteristic length representing the scale over
which the walls affect the piling of grains.

In this paper we are interested in numerical evaluation of the
applicability of the Janssen assumptions in the case of quasi 2-D
sandpiles formed by pouring granular material into thin vertical
Hele-Shaw cells. First we show that, similar to granular packs in
silos, the global pressure at the cell's bottom, $\sigma_b$,
saturates to a limiting value while the mass is increasing. We
then investigate the effect of wall-grain and grain-grain friction
coefficients on $\sigma_b$ and also global horizontal stress
$\sigma_v$ at the front and rear walls. An important issue is the
dependence of the Janssen coefficient $\kappa$ on system
parameters. We examine the effect of the friction coefficients
$\mu_p$ and $\mu_w$, as well as cell thickness $w$, on the value
of the Janssen coefficient  and observe interesting and sometimes
unexpected results.

\section{Model}
\begin{figure}
\begin{center}
\includegraphics[width=2.5 in  , height=3 in]{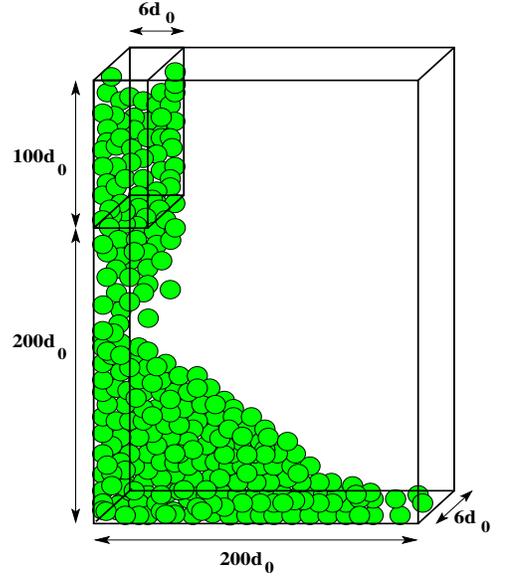}
\caption{A schematic view of the simulation box .}
\end{center}
\end {figure}
Our simulations are based on a Distinct
Element Method (DEM) scheme, originally proposed by Silbert et
al.~\cite{sil}. For the sake of clarity, we review the most
important aspects of it here. According to this model, the spheres
interact on contact through a linear spring dashpot interaction in
the normal and tangential directions to their lines of centers. In
a gravitational field $g$, the translational and rotational
motions of grain $i$ in a system at time $t$, caused by its
interactions with neighboring grains or walls, can be described by
Newton's second law, in terms of the total force and torque with the
following equations:
\begin{equation}
\textbf{F}^{tot}_{\textit{i}}=m_{\textit{i}}\textbf{g}+\sum\left(
\textbf{F}_{n_{\textit{i},\textit{j}}}+\textbf{F}_{t_{\textit{i},\textit{j}}}\right)
\end{equation}
\begin{equation}
\tau^{tot}_{\textit{i}}= -\frac{1}{2}\sum\textbf{r}_{\textit{i},
\textit{j}}\times\textbf{F}_{t_{\textit{i}, \textit{j}}}
\end{equation}
Where $m_{i}$ and $ \textbf{r}_{i,j}\equiv \textbf{r}_{i} - \textbf{r}_{j}$
are respectively, the mass of grain \textit{i} and the relative
distance between grains \textit{i} and \textit{j}. For two
contacting grains \textit{i}, \textit{j} at positions
$\textbf{r}_{i}$, $\textbf{r}_{j}$ with velocities
$\textbf{v}_{i}, \textbf{v}_{j}$ and angular velocities
$\omega_{i}, \omega_{j}$ the force on grain \textit{i} is computed
as follows. The normal compression $\delta_{i,j}$ is
\begin{equation}
\delta_{i,j}={d}-{r}_{\textit{i,j}}
\end{equation}
The relative normal velocity $\textbf{v}_{n_{i,j}}$ and relative
tangential velocity $\textbf{v}_{t_{i,j}}$ are given by
\begin{equation}
\textbf{v}_{n_{i,j}}=(\textbf{v}_{i,j}.\textbf{n}_{i,j})
\textbf{n}_{i,j}
\end{equation}
\begin{equation}
\textbf{v}_{t_{i,j}}=\textbf{v}_{i,j}-\textbf{v}_{n_{i,j}}-\frac{1}{2}(\omega_{i
}+\omega_{j}) \textbf{r}_{i,j}
\end{equation}
where $\textbf{n}_{i,j}=\frac{\textbf{r}_{i,j}}{r_{i,j}}$ with
$r_{i,j}=\vert \textbf{r}_{i,j}\vert$ and
$\textbf{v}_{i,j}=\textbf{v}_{i}-\textbf{v}_{j}$. The rate of
change of elastic tangential displacement $\textbf{u}_{t_{i,j}}$,
set to zero at the initial part of contact, is given by \cite{silp}
\begin{equation}
\frac{d\textbf{u}_{t_{i,j}}}{dt}=\textbf{v}_{t_{i,j}}-\frac{(\textbf{u}_{t_{i,j}
}.\textbf{v}_{i,j}) \textbf{r}_{i,j}}{r_{i,j}^{2}}
\end{equation}

The second term in Eq.(9) arises from the rigid body rotation
around the contact point and insures that $\textbf{u}_{t_{i,j}}$
always lies in the local tangent plan of contact. In Eqs. (1) and
(2), the normal and tangential forces acting on grain \textit{i}
are given by
\begin{equation}
\textbf{F}_{n_{i,j}}=f(\frac{\delta_{i,j}}{d})(k_n\delta_{i,j}\textbf{n}_{i,j}-
\gamma_{n}m_{eff}\textbf{v}_{n_{i,j}})
\end{equation}
and
\begin{equation}
\textbf{F}_{t_{i,j}}=f(\frac{\delta_{i,j}}{d})(k_{t}\textbf{u}_{t_{i,j}}-\gamma_
{t}m_{eff}\textbf{v}_{t_{i,j}})
\end{equation}
Where $k_{n,t}$ and $\gamma_{n,t}$ are elastic and viscoelastic
constants respectively and $m_{eff}=m_{i}m_{j}/(m_{i}+m_{j})$ is
the effective mass of spheres with masses $m_{i}$ and $m_{j}$. The
corresponding contact force on grain \textit{j} is simply given by
Newton’s third law, i.e., $\textbf{F}_{i,j}=-\textbf{F}_{j,i}$ .
For spheres of equal mass $\textit{m}$, as is the case in our
system, $m_{eff}=m/2$; $f(x=1)$ for the linear spring dashpot
(Hookian) model with viscoelastic damping between spheres
\cite{sil}.

Static friction is implemented by keeping track of the elastic
shear displacement throughout the lifetime of a contact. The
static yield criterion, characterized by a local grain friction
coefficient $\mu$, is modelled by truncating the magnitude of
$\textbf{u}_{t_{i,j}}$ as necessary to satisfy
$\vert\textbf{F}_{t_{i,j}}\vert<\vert\mu\textbf{F}_{n_{i,j}}\vert$.
Thus the contact surfaces are treated as “sticking” when
$\vert\textbf{F}_{t_{i,j}}\vert<\vert\mu\textbf{F}_{n_{i,j}}\vert$,
and as “slipping” when the yield criterion is satisfied
\cite{landry, Sun}.

In the following we present the results of our extensive molecular
dynamics (MD) simulations~\cite{md} in three dimensions on model
systems of $N$ mono-disperse, cohesionless and inelastic spheres
of diameter $d$ and mass $m$. The system is constrained by a
rectangular box with fixed rough walls and free top surface, as in fig1. A simulation was started with the
random generation of spheres without overlaps from top and left
corner of container, followed by a gravitational settling process
to form a stable heap. The results are given in non-dimensional
quantities by defining the following normalization parameters:
distance, time, velocity, forces, elastic constants, and stress
are, respectively measured in units of \textit{d},
$t_{0}=\sqrt{d/g}$, $v_{0}=\sqrt{dg}$, $F_{0}=mg$, $k_{0}=mg/d$,
and $\sigma=mg/d^2$. All data was taken after the system had
reached the steady state. Because of the complexity of the model,
there are a wide range of parameters that affect the results of
computation. We usually investigate the effect of a
single parameter varying in a certain range while other variables
are fixed to their base values as listed in Table \ref{table1}.

All the cases were simulated in three dimensions using a molecular
dynamics code for granular materials LAMMPS \cite{sil, Lammps}.
The equations of motion for the translational and rotational
degrees of freedom are integrated with either a third order Gear
predictor-corrector or velocity-Verlet scheme \cite{Plimpton}.

\section{saturation of wall normal stresses with mass}
We start from an empty vertical cell composed of two parallel
plates (front and rear walls) separated by an spacer with an
adjustable width $w$. A granular heap is formed in the cell by
pouring the grains on the bottom plate. These grains are released
from  a small box located on the top-left of the cell (see fig.1
and fig.2) where they have been randomly located in the first
place. The number of grains filling the small box, depends on the
material density and  the size of box. The rate of material
insertion $R$ can be varied by changing the box size. The process
repeated successively. A pile with well defined shape starts to
form as the number of the added grains grows. The angle of repose
can then be determined from the surface profile of the heap. The
angle of repose start from zero and reaches very soon a stable
value. Previous studies have demonstrated that the angle remains
constant with further insertion of grains~\cite{maleki}.
\begin{table}
\caption{\label{table1}Basic computational parameters}
\begin{ruledtabular}
\begin{tabular}{llll}
\\
Parameters&{\rm values}\\ \hline \\
\verb"Maximum Number of grains "$(N)$&$40,000$\\
\verb"grain-grain friction coef. " $(\mu_{p})$&$0.5$\\
\verb"wall-grain friction coef. "$(\mu_{w}) $&$0.5$\\
\verb"grain normal stiffness coef. "$(k_n )$&$2\times10^3(k_0)$\\
\verb"grain tangential stiffness coef. "$(k_t )$&$2/7 k_n$\\
\verb"grain normal damping coef. "$(\gamma_n )$&$50/(t_0)$\\
\verb"grain tangential damping coef. "$(\gamma_t )$&$50/(t_0))$\\
\verb"wall normal stiffness coef. "$(k_n )$&$2\times10^3(k_0)$\\
\verb"wall tangential stiffness coef. "$(k_t )$&$2/7 k_n$ \\
\verb"wall normal damping coef. "$(\gamma_n )$&$50/(t_0) $ \\
\verb"wall normal damping coef. "$(\gamma_t)$&$50/(t_0) $ \\
\verb"Time step increment "&$2\times10^{-3}$\\
\end{tabular}
\end{ruledtabular}
\end{table}

\begin{figure}
\begin{center}
\includegraphics[width=6 in  , height=3 in]{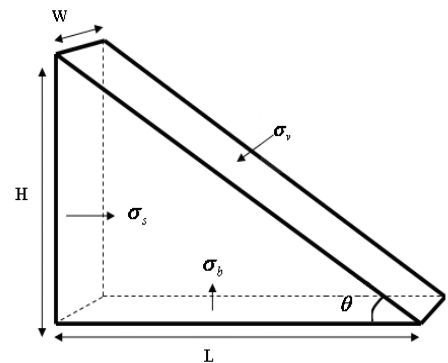}
\caption{The geometry of the pile and the wall normal stresses. }
\end{center}
\end {figure}

\begin{figure}
\begin{center}
\includegraphics[width=4 in  , height=2.8 in]{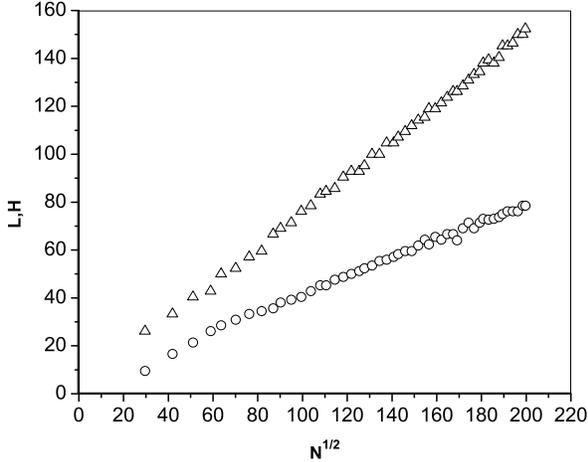}
\caption{ Variation of the Length $L$ (triangles) and the height $H$
(circles) of the sandpile with the square root of grain number
$N$ in a cell of thickness $w=6$}
\end{center}
\end {figure}
\begin{figure}
\begin{center}
\includegraphics[width=4 in  , height=2.8 in]{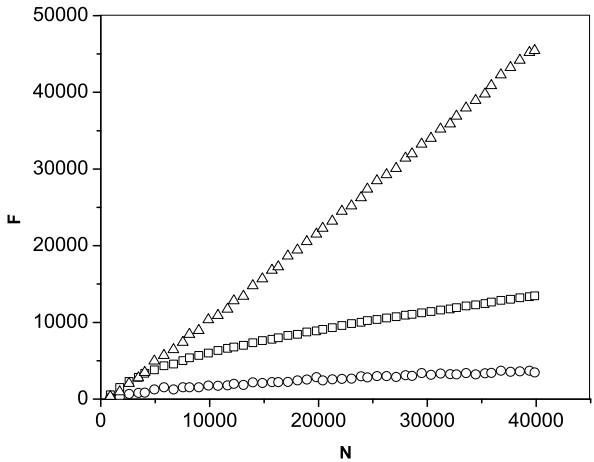}
\caption{ Variation of front or rear wall normal force, $F_v$
(triangles), apparent weight, $F_b$ (squares), and normal force on
the spacer, $F_s$ (circles) with grain number $N$ in a cell of
thickness $w=6$.}
\end{center}
\end{figure}
\begin{figure}
\begin{center}
\includegraphics[width=4 in  , height=2.8 in]{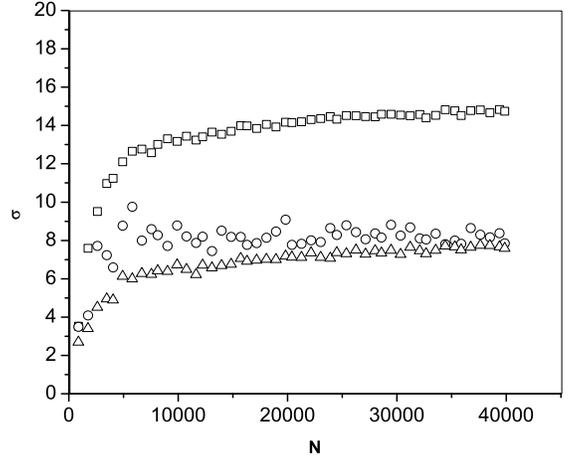}
\caption{ Variation of the front (or rear) wall normal stress $\sigma_v$
(triangles), bottom pressure $\sigma_b$ (squares), and normal
stress on the spacer $\sigma_s$ (circles) with grain number $N$ in
a cell of thickness $w=6$.}
\end{center}
\end {figure}
\begin{figure}
\begin{center}
\includegraphics[width=4 in  , height=2.8 in]{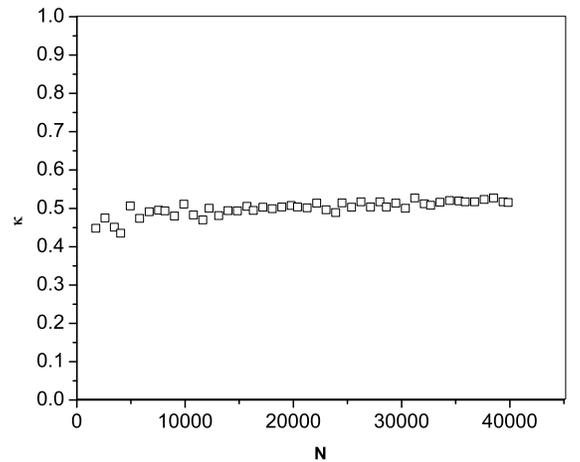}
\caption{ Estimation of Janssen coefficient $\kappa=\sigma_v /
\sigma_b$ from the data presented in fig.5 in a cell of thickness
$w=6$ with $\mu_p=\mu_w=0.5$.}
\end{center}
\end {figure}
The growth of  the height of the pile $H$ and length $L$
(measured from the pile profile) with grain number $N$ in a
typical cell with size $w=6$ is shown in fig.3. All the
other parameters are as those listed in Table \ref{table1}.
Clearly, the total pile mass $M$, is proportional to the number of
grains in the cell. It can be seen from fig.3 that after a short initial time, both $L$ and $H$ grow linearly with $\sqrt{N}$.
This is the expected behavior of a pile with constant density
$\rho$ and fixed angle of repose $\theta$.

The total normal forces exerted by the front (or rear) wall,
$F_v$, the spacer, $F_s$, and also the apparent weight of the
grain pile, i.e. the normal force exerted by the  bottom plate,
$F_b$, are all increasing functions of mass as depicted in fig.4.
Further investigations of these curves suggest that while $F_v$
grows linearly with $N$, both $F_b$ and $F_s$  increase linearly
with $\sqrt{N}$. Given the fact that except for the thickness $w$
the other pile  linear sizes are growing as $\sqrt{N}$, one can
deduce that all the global normal stresses $\sigma
_b=\frac{F_b}{Lw}$ (bottom pressure), $\sigma_s=\frac{F_s}{Hw}$
and $\sigma_v=\frac{2F_v}{LH}$ (horizontal stress) should approach
constant values after the transition period, that is when a
well-defined heap has been formed (see fig.5). From the data
presented in fig.5 one can see that compared to $\sigma_v$, the
value of $\sigma_s$ is larger at the beginning of the pile
formation process. In other words, in small heaps the horizontal
stress in the pile is not isotropic. However, as anticipated for
very large values of $N$, $\sigma_s$ eventually becomes almost
equivalent to $\sigma_v$, as in to the case of 3-D granular
columns.

From the obtained values for normal stresses, we have calculated
$\kappa$, the fraction of bottom pressure $\sigma_b$ which is
taken up by the front and rear walls, $\sigma_v$, as a function of
heap mass. The results presented in fig.6 demonstrate that ,
starting from smaller values, the Janssen coefficient $\kappa$
reaches very quickly to a maximal final value during the filling
process. This is in accordance with the Janssen theory which
assumes a constant value for $\kappa$ independent of the mass of
the injected material. Using the evaluated values of $\kappa$ in
the flat region of the curve, we have estimated $\kappa\cong
0.51\pm 0.01$ for this very case.

\section{effect of friction coefficients on wall normal stresses}
\begin{figure}
\begin{center}
\includegraphics[width=4 in  , height=2.8 in]{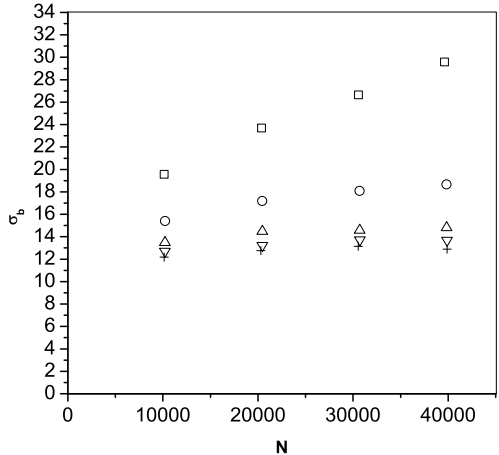}
\caption{  Variation of bottom pressure $\sigma_b$  with grain
number $N$ in a cell of thickness $w=6$ for different values of
$\mu_w$: $\mu_w=0.1$ ($\Box$); $\mu_w=0.3$ ($O$); $\mu_w=0.5$
($\triangle$); $\mu_w=0.7$ ($\nabla$); $\mu_w=0.9$ ($+$).}
\end{center}
\end{figure}

\begin{figure}
\begin{center}
\includegraphics[width=4 in  , height=2.8 in]{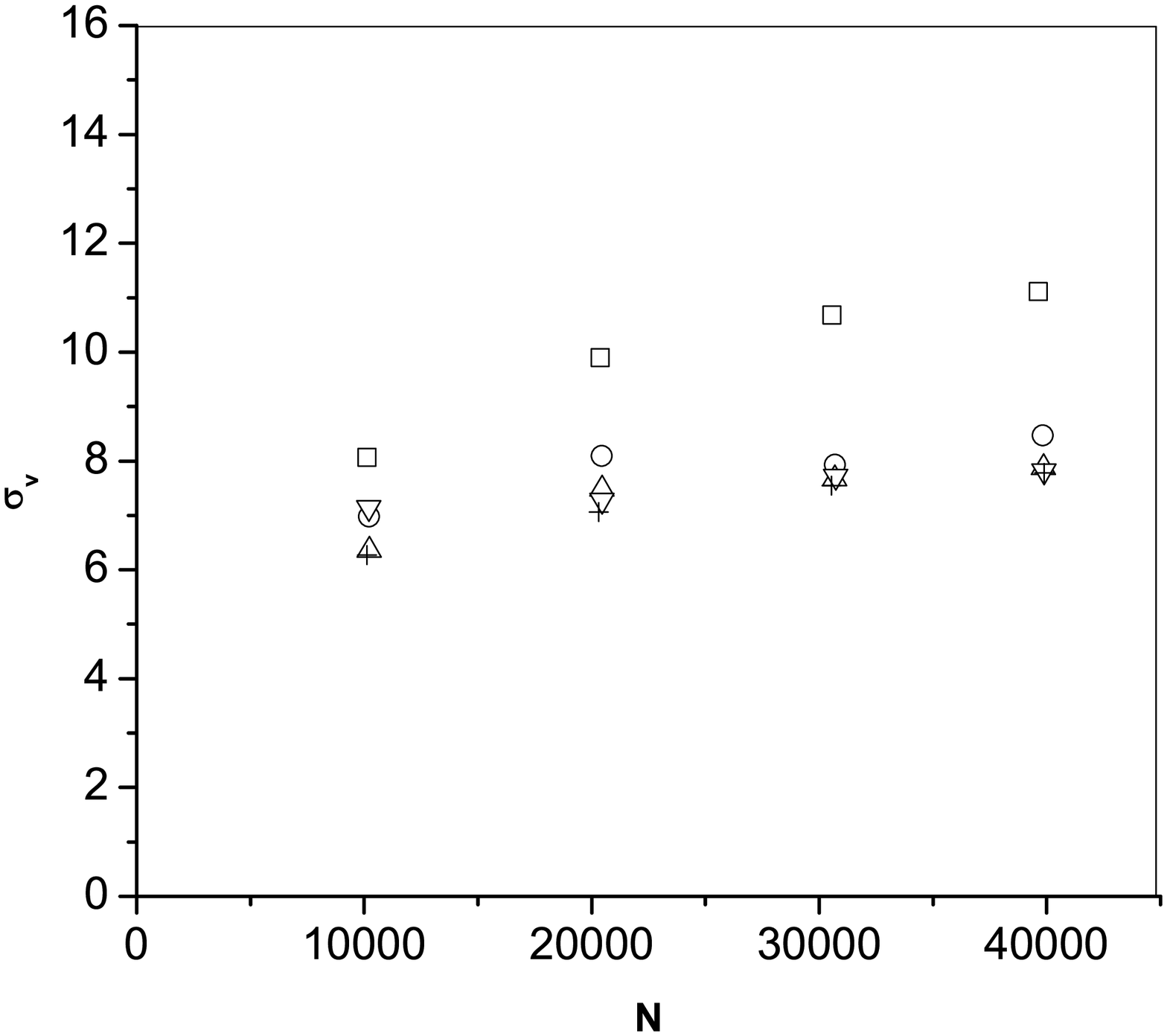}
\caption{  Variation of horizontal $\sigma_v$  with grain number
$N$ in a cell of thickness $w=6$ for different values of $\mu_w$:
$\mu_w=0.1$ ($\Box$); $\mu_w=0.3$ ($O$); $\mu_w=0.5$
($\triangle$); $\mu_w=0.7$ ($\nabla$); $\mu_w=0.9$ ($+$).}
\end{center}
\end {figure}

\begin{figure}
\begin{center}
\includegraphics[width=4 in  , height=2.8 in]{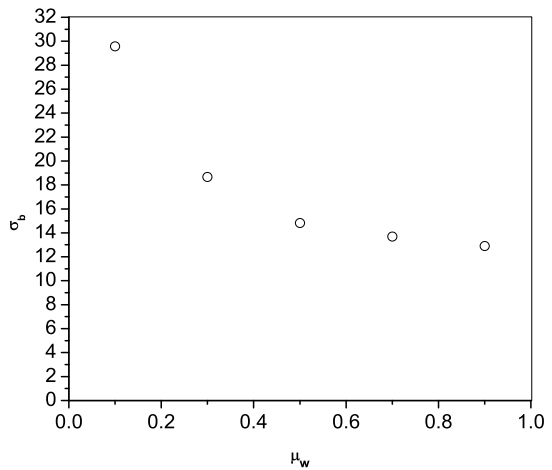}
\caption{ Variation of bottom pressure $\sigma_b$  versus $\mu_w$
for $ N=40000$  in a cell of thickness $w=6$, with $\mu_p=0.5$.}
\end{center}
\end {figure}

\begin{figure}
\begin{center}
\includegraphics[width=4 in  , height=2.8 in]{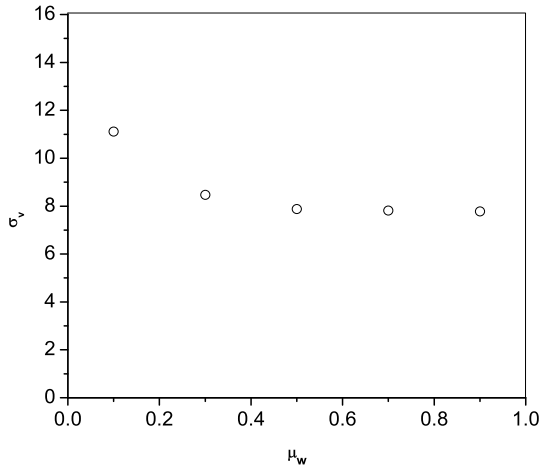}
\caption{ Variation of  horizontal stress $\sigma_v$ versus
$\mu_w$ for $ N=40000$,  in a cell of thickness $w=6$, with
$\mu_p=0.5$}
\end{center}
\end {figure}

\begin{figure}
\begin{center}
\includegraphics[width=4 in  , height=2.8 in]{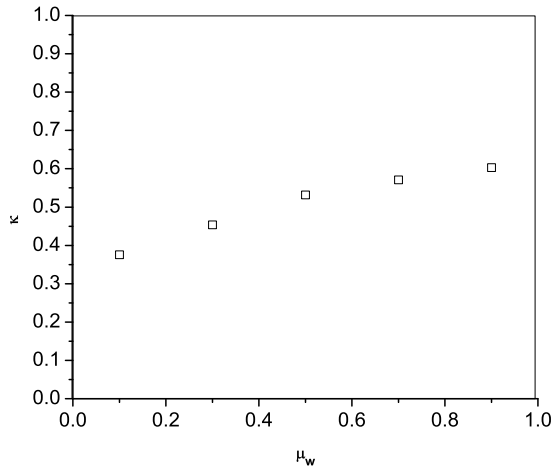}
\caption{ The dependence of  $\kappa$ on $\mu_w$ in a cell of
thickness $w=6$, with $\mu_p=0.5$.}
\end{center}
\end {figure}

The difference between real and apparent weight in granular packs
and confined sandpiles stems from the ability of the medium to
redirect part of the normal stress experienced at the bottom of
the container to the vertical walls surrounded it. For this we
need a stable network of grains in contact with each other which
spreads the forces through at the medium. Obviously, the
frictional forces plays the central role here, as they sustain the
stability of the contact networks and heap morphology.

Generally, the contribution of side walls, i.e. the shear stress
they impose on the confined granular material depends on the
strength of wall-grain frictional forces which is determined by
$\mu_{w}$, the wall-grain coefficient of friction. To examine the
behavior of the system when $\mu_w$ changes, we performed a series
of simulations on different heaps all formed in a cell of
thickness $w=6$ but with different wall friction coefficient
$\mu_w$. All the other parameters including grain-grain friction
coefficient have been fixed to the basic values listed in
Table\ref{table1}. From the data we have calculated  the variation
of normal stress $\sigma_b$, and  the behavior of the horizontal
stress $\sigma_v$, with the grain number $N$. For more clarity we
have also sketched in separate sheets the values of $\sigma_b$ and
$\sigma_v$ as a function of $\mu_w$, for the maximum grain number
we used in our simulations, $ N=40000$ (see fig.9 and fig.10).
Inspection of these curves shows that the wall normal stresses
almost cease to change with $\mu_{w}$ when it passes the value of
$\mu_w=0.5$, reach to their final minimal values. In other words,
for a granular material with specified grain parameters, there is
an upper limit to the contribution of walls in carrying the weight
of material. The case of $\mu_p=0.1$ is an exception. The
frictional forces between the container (and specially the
friction applied by the bottom)  are too small and as such the
formation of a heap with steady angle of repose needs more times
(grains).

From the above data we are also able to make an almost precise
estimate of the dependence of  $\kappa$ on $\mu_w$ by calculating
the value of $\sigma_v/\sigma_b$  for different values of $\mu_w$
(fig.11). Keeping track of variation of $\kappa$ during the
filling process we find out that the fraction of bottom pressure
converted to horizontal stress is almost constant, and the
variation of $\kappa$ with mass always follows a trend like that
of fig.6. Therefore, the estimated values for Janssen coefficient,
presented in fig.11 are precise even for the case of $\mu_w=0.1$.
This figure also suggests that $\kappa$ grows  with $\mu_w$ but
the growth slows down at larger values of $\mu_w$.

It is believed that one of the most effective medium parameters at
the microscopic level is the grain-grain friction coefficient
$\mu_p$~\cite{ned}. Therefore, we performed a series of
simulations for some selected values of $\mu_p$  while the other
parameters have been kept constant. The obtained  values for wall
normal stresses $\sigma_b$ and $\sigma_v$ in a pile consisting of
at most $N=40000$ grains have been presented in fig.12 and fig.13
respectively. Comparison of fig.12 with fig.9 shows that
increasing each of the friction coefficients ($\mu_p$ or $\mu_w$)
decreases $\sigma_b$ to a parameter dependent asymptotic value.
However, side-wall normal stress shows an opposite trend. As can
be seen from fig.13, $\sigma_v$ increases with $\mu_p$ before it
eventually saturates to a final value at larger $\mu_p $.
Considering the fact that the plie angle of repose $\theta$ grows
with both $\mu_p$ and $\mu_w$, one can deduce that both wall-grain
and grain-grain frictional forces increase the stability of the
sandpile but in two different ways. Increasing $\mu_w$ while
$\mu_p$ is kept constant, enhances total wall frictional forces as
the maximum wall friction $\mu_w F_N$ ($F_N$ can be each of the
wall normal forces $F_b$, $F_s$, or $F_v$) might increase even if
$\sigma_v$ decreases. On the other hand, by increasing grain-grain
friction coefficient, the horizontal normal stresses increase, as
suggested by fig.13. These also leads to the strengthening of the
maximum frictional force experienced at each side walls.

Of particular interest is the way friction coefficients affect the
behavior of the Janssen coefficient. We have chosen four different
values for $\mu_w: 0.05,0.3, 0.5$ and $1.0$, and calculated the
variation of $\kappa$ with $\mu_p$ for each case. The results are
summarized in a unique diagram in fig.14. From this figure we find
that putting aside the case $\mu_p=0.05$, the value of $\kappa$
starts to grow rather rapidly at small values of $\mu_p$, to a
final maximal value which is itself an increasing function of
$\mu_w$  in all curves. Our simulations also show that for the
case $\mu_p=0.05$ the value of $\kappa$ is almost insensitive to
the variation of wall-particle friction coefficient. This might be
assigned to the isostaticity of the granular packing when $\mu_p$
is too small, similar to the phenomenon observed in Ref.7.
Besides, the saturation of $\kappa$ with $\mu_p$ does not show a
simple trend: while the flat part of the first three curves
happens to start at $\mu_p\geq \mu_w$, it begins at a value of
$\mu_p$ much smaller than $\mu_w$ for the fourth curve
($\mu_p=1.$). This observation resembles to the results of Ref.7
too, which demonstrate the force ambiguity does not follow a
monotonic variation with $\mu_p$.

It should be noted that for a more comprehensive study of the
effect of friction coefficients on $\kappa$ we need to extend the
range of variation of $\mu_p$ and $\mu_w$ on the logarithmic
scale. But the problem is that for small values of $\kappa$ a heap
hardly formed. In fact, a more suitable system for such
experiments is a granular pack formed in a silo instead of a
granular heap. Because the formation of a well defined pack is
always guaranteed even if the frictional forces are not
significant.

\begin{figure}
\includegraphics[width=4 in  , height=2.8 in]{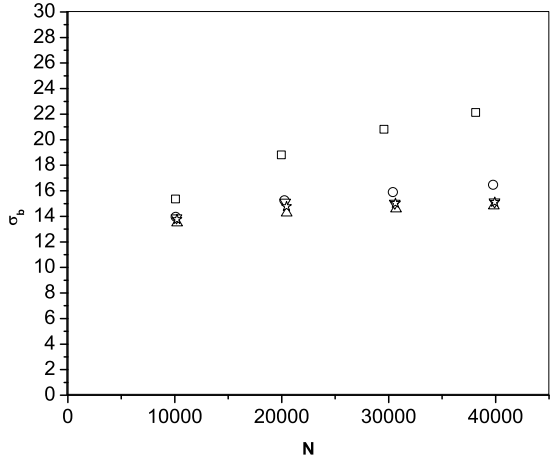}
\caption{  Variation of bottom pressure $\sigma_b$  with grain
number $N$ in a cell of thickness $w=6$ for different values of
$\mu_p$: $\mu_p=0.1$ ($\Box$); $\mu_p=0.3$ ($O$); $\mu_p=0.5$
($\triangle$); $\mu_p=0.7$ ($\nabla$); $\mu_p=0.9$ ($+$).}
\end{figure}

\begin{figure}
\begin{center}
\includegraphics[width=4 in  , height=2.8 in]{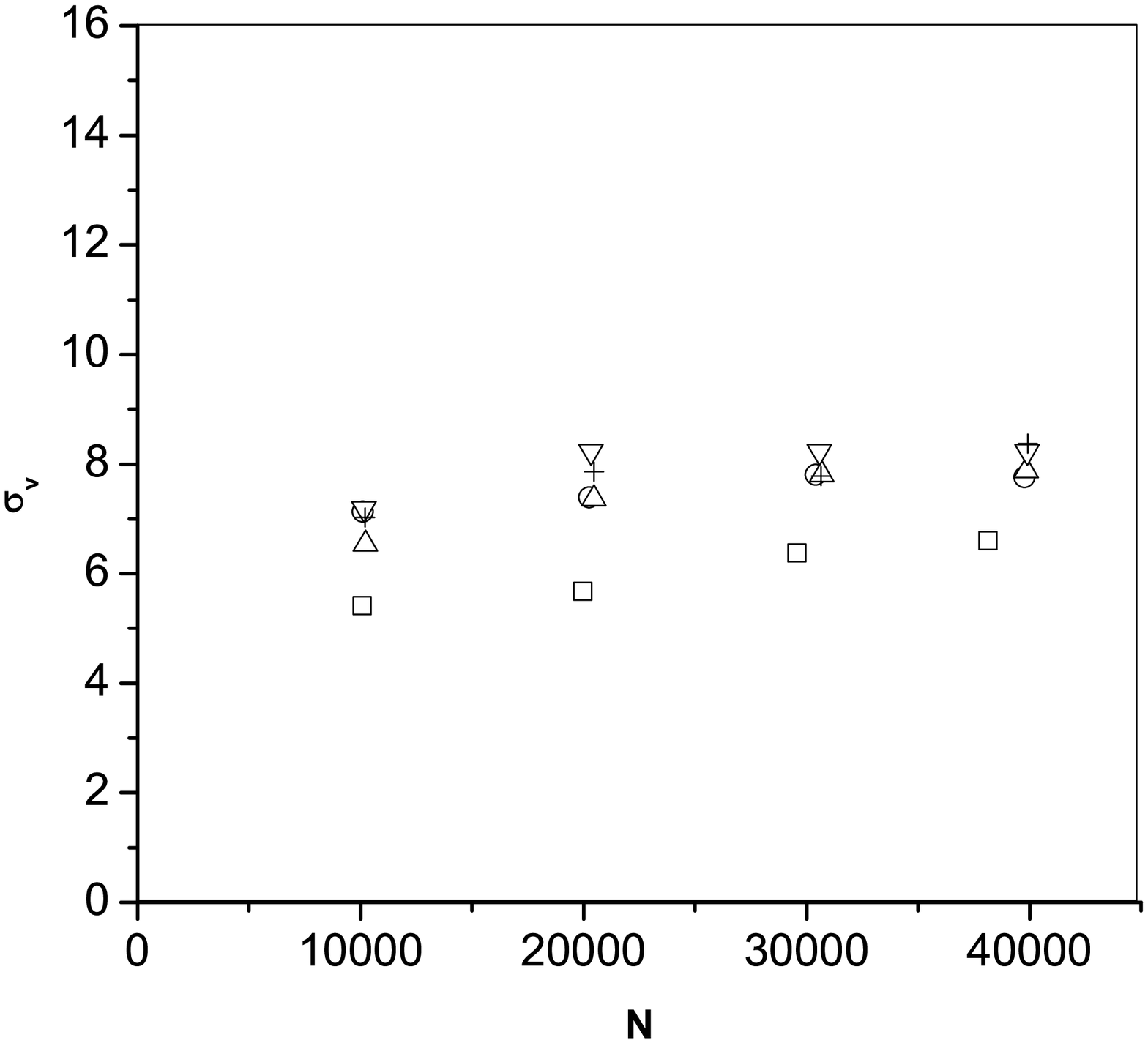}
\caption{  Variation of horizontal stress $\sigma_v$  with grain
number $N$ in a cell of thickness $w=6$ for different values of
$\mu_p$: $\mu_p=0.1$ ($\Box$); $\mu_p=0.3$ ($O$); $\mu_p=0.5$
($\triangle$); $\mu_p=0.7$ ($\nabla$); $\mu_p=0.9$ ($+$).}
\end{center}
\end {figure}

\begin{figure}
\begin{center}
\includegraphics[width=3 in  , height=2.5 in]{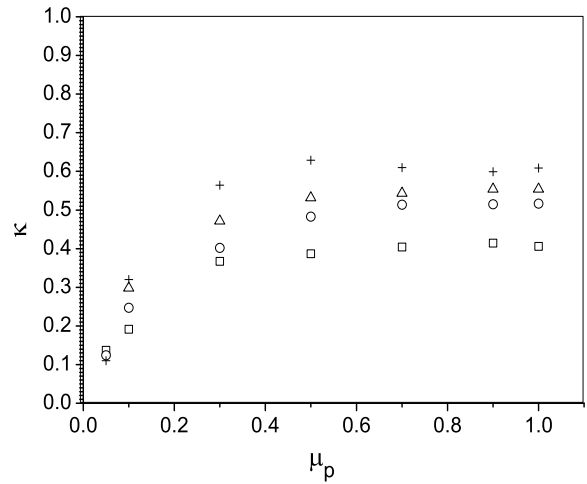}
\caption{ The dependence of  Janssen coefficient $\kappa$ on
grain-grain friction coefficient $\mu_p$ for different values of
$\mu_w$: $\mu_w=0.05$ ($\Box$); $\mu_w=0.3$ ($O$); $\mu_w=0.5$
($\triangle$);$\mu_w=1.0$ ($+$).}
\end{center}
\end {figure}
\section{Approaching to 3-D sandpiles}

The effect of the cell thickness $w$ on the stability of a
confined sandpile has been the subject of both experimental and
numerical studies~\cite{gh,bol,xy1,xy2,cou,maleki}. Careful
experiments and measurements reveal that the presence of front and
rear walls increases greatly the angle of repose $\theta$ as well
as the maximal angle of movement $\theta_m$ (the maximal angle
that a granular medium can reach when carefully tilted). These
observations show that by increasing the cell thickness, the angle
of repose, $\theta(w)$, decreases and it eventually settles down
to an asymptotic value $\theta_\infty$ when the cell thickness
becomes very large compared to the size of grains. The observed
phenomenon is universal and does not depend on the formation
history of the pile.

To investigate the effect of $w$ on the wall normal stresses, we
conducted a series of simulations on cells with different
thicknesses, where all the other system parameters ad boundary
conditions are as listed in Table1. The results for $\sigma_b$ and
$\sigma_v$ are presented in fig15 and fig.16. These curves
indicate again that normal stresses are going to  saturate with
mass for all the cell thickness $w$ but with different rates. An
interesting feature is that although  $\sigma_b$ grows with cell
thickness at smaller values of $w$, but the grows stops when $w$
becomes large compared to a single grain size. Meanwhile, the
behavior of normal stresses at lateral walls is different. In fact
the values of $\sigma_v$ and $\sigma_s$ (not shown) decrease with
$w$ at smaller cell thickness but again it seems both of them
become almost $w$ independent when $w$ is large enough. These
behaviors are very similar to the variation of $\theta$ with
$w$~\cite{gh,xy1,xy2,cou,maleki}. In both cases there seems to be
a transition from quasi two dimensional systems when the cell is
just few grain diameters across, to three dimensional case where
the thickness of the container is much greater than a single grain
size.

Let us consider now the behavior of $\kappa$, the the fraction of
bottom stress $\sigma_b$ which is redirected to each of the side
walls, with the cell thickness grows to higher values. In fig.17
we have presented our calculation of $\kappa$ and $\theta$ for
cells with different $w$, all of them contain $40000$ grains. The
other parameters are the basic values listed in Table1. As the
figure shows,  for small values of $w$ the major fraction of
$\sigma_b$ is converted to side-wall normal stresses making the
contribution of frictional forces more pronounced at these small
scales. However when the spacing between front and rear walls
become large, compared with the size of constituent grains,
$\kappa$ and consequently  the normal stress exerted by front or
rear wall approaches constant values. In fact, the behavior of
$\kappa$ is very similar to the way $\theta$ varies with $w$. This
is a new concept and should be considered in physical modelling of
wall effects on the angle of repose of confined heaps ( as well as
angle of movement), especially when the spacing between front and
rear walls is of the order of few grain diameters.
\begin{figure}
\begin{center}
\includegraphics[width=4 in , height=2.8 in]{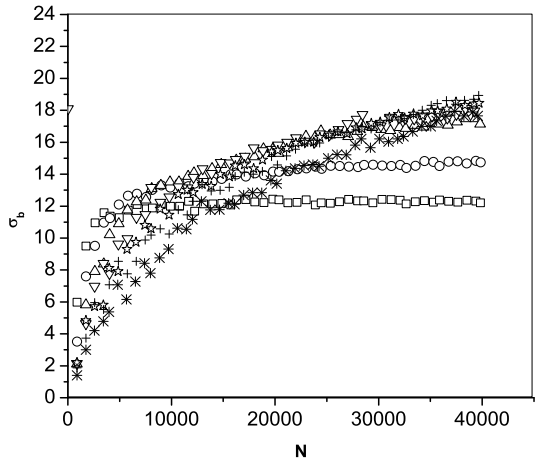}
\caption{ The variation of bottom pressure $\sigma_b$ with grain
number $N$ for various cell thickness $w$: $w=4 (\Box)$; $w=6
(O)$; $w=8 (\triangle)$; $w=10 (\nabla$); $w=12 (\star)$; $w=14
(+)$; $w=18 (\ast)$}
\end{center}
\end{figure}

\begin{figure}
\begin{center}
\includegraphics[width=4.5 in  , height=2.8 in]{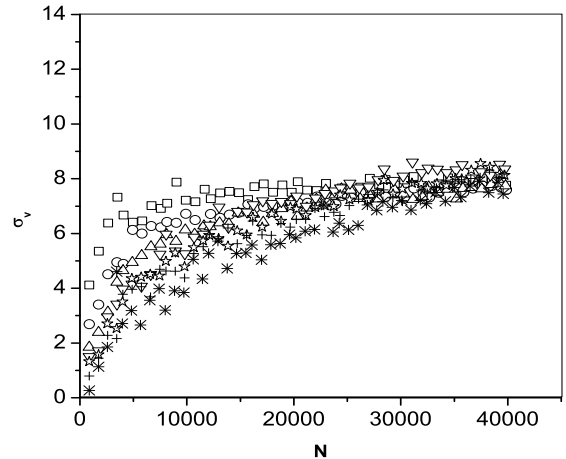}
\caption{ The variation of front-rear stress, $\sigma_v$ with
grain number $N$ for various cell thickness $w$: $w=4 (\Box)$;
$w=6 (O)$; $w=8 (\triangle)$; $w=10 (\nabla$); $w=12 (\star)$;
$w=14 (+)$; $w=18 (\ast)$}
\end{center}
\end{figure}

\begin{figure}
\begin{center}
\includegraphics[width=3in  , height=2.3 in]{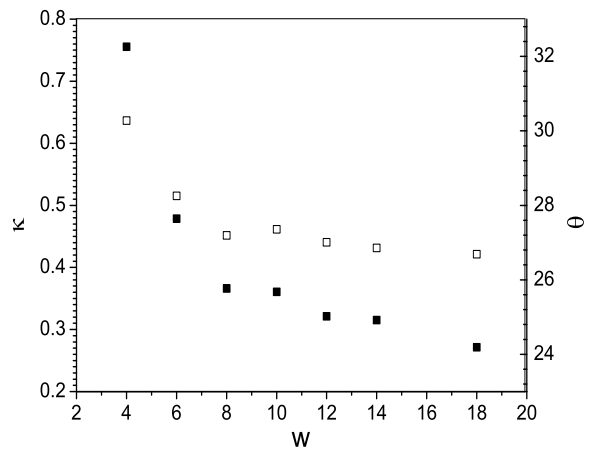}
\caption{ Dependence of  Janssen coefficient $\kappa$ (squares)
and angle of repose $\theta$ (filled squares)on the cell thickness
$w$ with $\mu_p=\mu_w=0.5$.} \label{fig12}
\end{center}
\end {figure}

\section{CONCLUSIONS}

We have used 3-D MD simulations to investigate the effects of a
container's walls on the stability of  sandpiles formed by pouring
dried spherical grains into a container with small separation
between front and rear walls. We observed the Janssen's effect,
namely the saturation of global pressure at the bottom of the
container, although the geometry is different from the traditional
silos where the standard Janssen analysis is applied. The same
behavior has been observed for front or rear wall's normal
stresses and the vertical slab supporting the pile. Increasing the
size of the container we witnessed the saturation effects again
but at larger number of grains. We saw that for larger cell
thickness, the saturation of normal stresses happened at larger
values of mass, with stress at the bottom plate of the container
increases with $w$ and the opposite trend for $\sigma_v$.

We have also studied the dependence of the wall normal stresses
with wall friction coefficient $\mu_w$ and with grain friction
coefficient $\mu_p$ and found that increasing both of these
parameters causes a reduction in normal stress felt at the cell's
bottom.  Meanwhile, we saw that the mechanisms of the observed
reduction in apparent weight while a friction coefficient is
increasing, depends on which of $\mu_p$ and $\mu_w$ is subject to
change. While the results of simulations show that by increasing
wall-grain friction coefficient $\mu_w$, the horizontal stress
$\sigma_v$ approaches smaller values,  the changes are such that
the overall value of $\mu_w F_N$ which is the maximum of wall
friction becomes larger and as such a stable heap with greater
$\theta$ forms. On the other hand, when $\mu_w$ is fixed and
$\mu_p$ grows, we  saw an increase in the calculated values of
$\sigma_v$, which means the wall friction limit $\mu_w F_N$, is
again growing with $\mu_p$.

We have estimated the Janssen's coefficient $\kappa$ and found
that not only the grain properties, but also the cell geometry and
wall- grain interaction affect this quantity. Our observations
suggest that kappa is affected by cell thickness $w$ when $w$ is
less than about $8d$ and therefore, the conventional assumption of
$\kappa$ being only dependent to the granular material microscopic
characteristics is under question at small $w$. We believe that
this is a new concept and should be considered in theoretical
treatment of granular material at small scales. The variation of
$\kappa$ with $\mu_p$ when other experimental parameters are kept
constant is also interesting. Our observations indicate that
$\kappa$ grows almost linearly with $\mu_p$ for small values but
reaches to an asymptotic value at larger values of $\mu_p$.
Meanwhile, when $\mu_p$ is not too small, increasing $\mu_w$
enhances the pressure transmission from bottom to the vertical
walls, making $\kappa$ an increasing function of $\mu_w$.

We are thankful to E.Nedaaee Oskoee for his help and J.Stivenson
for reading this manuscript carefully. We are also grateful to
IASBS parallel computer center for the computer facilities.

\end{document}